\documentclass{emulateapj}

\begin{document}

\title{HST/STIS Optical Transit Transmission Spectra of the hot-Jupiter HD209458\MakeLowercase{b}}

\author{David K. Sing\altaffilmark{1}}
\author{A. Vidal-Madjar\altaffilmark{1}} 
\author{J.-M. D\'{e}sert\altaffilmark{1}}
\author{A. Lecavelier des Etangs\altaffilmark{1}}
\author{G. Ballester\altaffilmark{2}}

\altaffiltext{1}{Institut d'Astrophysique de Paris, CNRS; Universit\'{e} 
Pierre et Marie Curie, 98 bis bv Arago, F- 75014 Paris, France; sing@iap.fr}

\altaffiltext{2}{Lunar and Planetary Laboratory, University of Arizona, 
Sonett Space Science Building, Tucson, AZ 85721-0063, USA}

\submitted{Accepted to Ap.J. May 6, 2008}

\begin{abstract}

We present the transmission spectra of the hot-Jupiter HD209458b taken
with the Space Telescope Imaging Spectrograph aboard the Hubble Space
Telescope.  Our analysis combines data at two resolutions and applies
a complete pixel-by-pixel limb-darkening correction to fully reveal the spectral
line shapes of atmospheric absorption features.  
Terrestrial-based Na I and H I contamination 
are identified which mask the strong exoplanetary absorption signature in the Na core, which
we find reaches total absorption levels of $\sim$0.11\% in a 4.4{\AA} band.
The Na spectral line profile is characterized by a wide absorption profile at the lowest
absorption depths, and a sharp transition to a narrow absorption profile at higher
absorption values.  
The transmission spectra also shows the presence of an additional
absorber at $\sim$6,250{\AA}, observed at both medium and low resolutions.  
We performed various limb-darkening tests, including using high precision limb-darkening measurements of the sun 
to characterize a general trend of Atlas models to slightly overestimate 
the amount of limb-darkening at all wavelengths, likely due to 
the limitations of the model's one-dimensional nature.  We conclude 
that, despite these limitations, Atlas models can still successfully 
model limb-darkening in high signal-to-noise transits of solar-type stars, like HD209458, 
to a high level of precision over the entire optical regime (3,000-10,000 
{\AA}) at transit phases between 2$^{nd}$ and 3$^{rd}$ contact.  

\end{abstract}


\section{Introduction}
Transiting hot-Jupiter planets offer a tremendous opportunity to study
the characteristics of extrasolar planets.  During a transit event,
both the opaque body of the planet as well as its atmosphere blocks
light from the parent star.  A precise determination of the radius of
the planet can be made from the total obscuration, while partial
transmission of light through the exoplanet's atmosphere, along with
its wavelength dependence, allows for the detection of composition and
structure \citep{Charby,Ballester07}.  In the middle of the visible
spectrum, transits can probe atmospheric pressures from about 0.1 bar
to 1$\times10^{-6}$ bar, containing the troposphere (lower atmosphere), 
as well as the stratosphere and mesosphere (middle atmosphere).  One exoplanet in particular, HD209458b,
has been ideal for these studies and holds the distinction of the first detection of an
extrasolar planetary atmosphere \citep{Charby} as well as the discovery
of an escaping atmosphere \citep{Vidal02,Vidal03}.  These transit
studies were performed with the Hubble Space Telescope ({\it HST})
using the Space Telescope Imaging Spectrograph instrument ({\it STIS}).

Here we expand upon earlier {\it HST STIS}
analysis \citep{Brown01,Charby,Knut,Ballester07} in order to fully probe the line shape
of atmospheric Na in HD209458b, and quantify other absorbers in the transmission spectrum.  
We use {\it STIS} data obtained during planetary transit at two spectral resolutions
(low and medium).  The two datasets are combined to extend 
the measurements over the entire optical regime, providing a
way to simultaneously measure both the narrow core and wide line wing Na
absorption.  We describe these observations in \S 2, perform various limb-darkening tests
in \S 3, present analysis of this data in \S 4, and discuss the results in \S 5.  
From our resulting spectra, a detailed atmospheric model fit is given in \cite{Sing08}, which discusses condensation
of sodium, as well as the implications of the atmospheric temperatures found.  In addition, details of H$_{2}$
Rayleigh scattering are presented in \cite{Lecavelier08} while \cite{desert08} explores absorption by atmospheric TiO \& VO.

\section{Observations}
\subsection{Data reduction}

The HST STIS G750M, G750L, and G430L observations of HD209458b 
analyzed here are also detailed in \cite{Charby}, \cite{Ballester07}, 
\cite{Brown01}, and \cite{Knut}. The G750M data 
consist of 684 spectra covering the wavelength range of 5,518-6,382 
{\AA}, with a resolution R of $\lambda$/$\Delta\lambda=$5,540 ($\sim$2 pixels; 1.1 {\AA}), 
and taken with a wide 52''-2'' slit to minimize slit light losses. This observing technique 
produces spectra which are photometrically accurate near the 
Poisson limit during the transit event. Four transits were observed, 
although the first suffered from a database error of the position 
of the spectrum and, as in \cite{Charby}, was not used 
for this analysis.

The dataset was pipeline-reduced with the latest version of CALSTIS 
and cleaned for cosmic ray detections before performing spectral 
extractions. The aperture extraction was done in IRAF software 
with a 30-pixel-wide aperture, no background subtraction, and 
the optimal extraction algorithm of \cite{Horne86}. 
This extraction technique produces spectra with \ensuremath{\sim}5-6\% 
more flux than simply summing over the cross-dispersion direction, 
as the stellar counts from the wings of the PSF can be added 
without additional noise.

The spectra were Doppler corrected to the heliocentric rest frame, 
with the velocity shifts on the order of one pixel.
An additional sub-pixel shift in the dispersion direction 
was removed in each spectrum by cross-correlation against a template 
spectrum, which consisted of the mean pixel-shifted spectra. 
Such sub-pixel shifting are likely due to the jitter of the spacecraft which
changes the position of the star within the slit.  The shifts are
typically on the order of $\sim$0.1 pixel within a orbit (the typical guiding
accuracy of HST) and $\sim\pm$0.5 pixel from orbit to orbit. 
Several iterations of measuring pixel-shifts, applying those 
shifts, and producing a new template were performed in order 
to produce a reliable final template spectra for cross-correlation. 
With this method, we were able to align the spectra to \ensuremath{\sim}0.1 
pixel.  With different Doppler and pixels-shifts for each spectrum, the PSF of the stellar 
absorption line profiles can be reconstructed with 
a much greater sampling than can be obtained from an individual spectrum (see Fig. 1).

The extracted STIS spectra were then used to create a photometric 
time series covering the G750M bandpass by integrating the flux 
from each exposure. The resulting photometric light curve exhibits 
all the systematic instrumental effects noted in \cite{Brown01}. 
As in both \cite{Brown01} and \cite{Charby}, 
we corrected for these systematic effects by discarding the first 
exposure of each orbit, discarding the first orbit of each visit, 
and applying flux corrections by removing a fourth order polynomial 
fit to the photometric time series phased on the HST orbital 
period. The photometric corrections were then applied to the 
spectra themselves, such that different photometric bands could 
easily be produced.

The Na D wavelength region is also covered by HST STIS G750L 
observations, described by \cite{Knut}, though at much 
lower resolution. The G750L data set was also reduced, treated 
in the same manner as the G750M. The cross-correlation analysis, 
however, only was performed over the blue-half of the data, such 
that our measured pixel-shifts would be independent of the observed 
fringes.

\subsection{Telluric contamination}

Telluric contamination can be seen in a photometric G750M time-series 
of the Na core where we see a large variation of 4-5\% over an 
HST orbit (see Fig. 2). This variation is observed to be 
correlated with the HST orbital phase, and repeats in a similar 
manner within each orbit of a visit, and occurs in all three 
visits analyzed.  The contamination occurs for both Na lines over
the same radial velocities, $\sim$-5 to -20 km s$^{-1}$ from the Na D line cores, and is
consistent with the expected velocity of the earth during the HST observations.
Given the size of the variation, it is too large 
to be due to the known instrumental systematic effects, which 
only affect the light curves \ensuremath{\sim}0.1\% during an orbit. 
In addition, other strong stellar lines such as Fe I do not show 
any signs of a similar contamination, ruling out possible detector 
efficiency and count-rate dependent issues as the cause of the 
observed line core variations.  The contamination can also be 
seen directly in the G750M spectra (Fig. 1), observed as 
a \ensuremath{\sim}0.3{\AA} region in the sodium D cores with little-to-no 
flux drop during planetary transit, though the actual contaminating region could result
from a smaller unresolved region.  Absorption by interstellar 
Na could also be affecting the contaminated region although it's effect would be constant 
and should not vary with the HST orbital period.  The G750L data-set 
also shows telluric Na D contamination, with H$_{\alpha}$ showing a similar 
effect as well. With lower resolution, the telluric signature 
is seen in photometric time-series measurements of the line cores 
having, as expected, a correspondingly lower amplitude than the 
medium resolution data.

\subsection{Sodium core absorption signature}

We performed a double differential photometric measurement, similar 
to \cite{Charby}, selecting 23 wavelength bands from 
\ensuremath{\sim}4-100 {\AA} centered on the Na D lines. In the original 
analysis, \cite{Charby} used three different wavelength 
bands. We chose to use the sum of two bands, each centered on 
one component of the Na doublet for the two wavelength bands 
smaller than the Na D doublet separation. For each band, we integrated 
the spectral flux, producing photometric light curves. We then 
normalized each visit to their average out-of-transit flux. The 
two narrowest bands showed terrestrial contamination which modulate 
with the HST orbital period, which we removed by fitting each visit of the HST phased
out-of-transit flux with a fourth-order polynomial applying the fit to the in-transit orbits, as was done 
in a similar manner for the instrumental systematic effects. 
The photometric curves were then subtracted from a comparison 
band composed of the wavelength regions between 5,818-5,843 {\AA} 
and 5,943-5,968 {\AA}, picked to be identical to the ``wide'' comparison 
band in \cite{Charby}. The relative planetary Na absorption 
during transit was then computed from the mean of the difference 
light curve over the in-transit phases between second and third 
contact, with its associated error computed from the variance 
of the data. Changing the width of the comparison bands, in the 
same manner as \cite{Charby} produced similar results 
for corresponding bands. Our resulting absorption profile (see 
Fig. 3) matches the results of \cite{Charby} and, 
in addition, reveals the strong absorption from the Na D line 
centers. When left uncorrected for terrestrial-based contamination, 
the two narrowest bands produce similar absorption values, but 
with error bars large enough for the absorption to be consistent 
with a null detection.

A double differential spectroscopic measurement was also performed 
over the Na region. For the spectroscopic measurement, each spectra 
was first interpolated, using a cubic spline, onto a common wavelength 
scale. The spectral flux at each pixel was then normalized by 
the average out-of-transit flux, then subtracted from the average 
normalized flux of the ``wide'' comparison region.  The absorption 
at each wavelength during transit was then measured as performed for the 
photometric analysis, including telluric corrections for each pixel.  The resulting differential spectra shows 
two statistically significant positive absorption peaks at the 
Na D lines with each having a large negative absorption core where the  
the terrestrial contamination is largest (see Fig. 4).  
The telluric corrections on the Na lines have a modest affect on the absorption values outside
the two largest contaminated pixels, as the telluric perturbation originates in a small sub-pixel region, with
nearby pixels effected through the $\sim$2 pixel instrument resolution. 
Excluding the two highly contaminated pixels affectively removes the majority of the contamination.
In a 4 pixel bin, the average value of the two line cores is 
consistent with the 4.4 {\AA} band photometry measurement indicating that both approaches to correcting the 
contamination give consistent values.  
In our final limb-darkened corrected transmission spectra (see \S4.3), we
ultimately choose to apply the telluric corrections over the Na region
while also leaving out the two highest contaminated pixels.  
Binning the differential spectra by 3 pixels (1.66 {\AA}) shows the Na D2 line is stronger than the absorption of 
the D1 line with absorption values 0.095$\pm$0.02\% and 0.051$\pm$0.02\% respectively, 
and an absorption ratio of D2/D1 = 1.9$\pm$0.9.  
The differential photometric measurement
includes these two pixels as their effects are negligible at large band widths and correctable 
for the 4.4 and 8.9 {\AA} bands.

\subsection{Limb-darkening corrections}

Correcting for limb-darkening effects allows a full transmission 
spectrum to be produced over the entire wavelength range of the 
STIS data. Similar to \cite{Ballester07}, we computed nonlinear 
limb-darkening coefficients to estimate and correct for the limb-darkening 
effects in the transit light curve. Though other limb darkening 
laws could in principle be used (for example: linear, quadratic, 
square-root, or logarithmic), \cite{Claret00} found 
the nonlinear law best represents the intensity distributions 
calculated from 1-dimensional plane parallel atmospheric models 
used here.

We choose a Kurucz Atlas stellar atmospheric model\footnote{http://kurucz.harvard.edu/stars/hd209458} 
based on the 
model's limb-darkening performance, performance vs. other 1D 
models, model availability, and development within the literature. 
Limb-darkening calculated from Phoenix models (Claret 2000) were 
found to predict a stronger limb-darkening, compared to Atlas 
models, by a few percent in the photometric bands B, V, R, I, 
J, H and H, but up to 40\% weaker in U and u' bands. As the Atlas 
models performed better compared to the sun (see \S 3.2), 
they were chosen over Phoenix models for our study. 

Most studies fitting transit light curves have either chosen 
to adopt an Atlas stellar atmospheric model to account for limb-darkening 
effects, or fit for limb-darkening parameters assuming a limb-darkening 
law. Knutson et al. (2007), who also used an Atlas model for 
HD209458, found similar planetary parameters when comparing the 
two methods, suggesting that model based limb-darkening fits 
can perform well. Knutson et al. (2007) chose to use the limb-darkening 
calculated from the Atlas models in their results, due to the 
more precise planetary parameters found, and concerns that fit 
limb-darkening coefficients could be affected by residual correlation 
in the light-curve. Such concerns are valid, as we found residual 
correlated ``red'' noise ultimately limits the maximum signal-to-noise 
S/N achievable in this study (see \S 4.2). 
In addition, as our limb-darkening corrections can be on a pixel-by-pixel 
level, a light-curve generated by any one particular pixel will 
not have sufficient S/N to accurately fit for limb-darkening 
coefficients.

The four nonlinear coefficients were computed using intensity 
spectra from a Kurucz 1-dimensional plane parallel model atmosphere 
calculated at 20 different angles spread uniformly in \textit{\ensuremath{\mu}}, 
fitting for the coefficients in the desired wavelength range. 
We calculated coefficients for the wavelengths ranges used in 
the G750M differential measurements, and each pixel of the G750M 
and G750L dispersions. Using the fit coefficients, we then used 
the theoretical transits of Mandel \& Agol (2002) to determine 
and correct for limb-darkening effects. The best-fit system parameters 
of Knutson et al. (2007) were used to calculate the impact parameter 
and planetary/star radius contrast. For the differential photometric 
measurement probing the Na line cores, designed by Charbonneau et al.
(2002) to be largely limb-darkened independent, all the corrections 
had a less than 1$\sigma$ effect with corrections to our narrowest band 
only increasing the absorption by 0.005\% (see Fig. 3).

\section{Limb-Darkening Tests}

\subsection{Stellar line vs. continuum limb-darkening strength measurement}

When measuring Na in the G750M photometric bandpass, limb-darkening 
effects increase with narrowing Na bands. Specifically, the stellar 
limb-darkening is less in the Na line core compared to the surrounding 
continuum, producing transit light-curves which are slightly 
more ``box'' shaped with a depth at mid-transit slightly smaller. 
When comparing the transit light-curves of narrow regions centered 
on the Na line to that of a surrounding continuum, with the intention 
of measuring a difference in apparent planetary radius, the overall 
effect is to inherently underestimate the signature. Using the 
G750M spectra, we measured the magnitude of this effect using 
the stellar absorption lines (Fe I, Ca I, Si I, etc.) and the 
continuum. 
No other large planetary atmospheric signature other than 
Na was detected in the stellar lines of the G750M data (Charbonneau 
et al. 2002), making them excellent probes of the effects of 
limb-darkening independent of differing planetary radii. We produced 
two separate photometric light curves, one that contained only 
the flux from stellar continuum and one that contained only flux 
from the stellar absorption lines. The Na region was not included 
in either photometry. We then normalized the two curves by their 
out-of-transit flux and took their difference. The average in-transit 
difference was measured, as in the differential Na analysis, 
to be -0.0044\ensuremath{\pm}0.0015\%. Using the model atmosphere and 
calculating theoretical transits for the selected wavelengths, 
we found a theoretical difference of -0.0038\% which matches 
the observed value. This confirms that: (1) the limb-darkening 
in the stellar lines is smaller than the continuum, (2) the effect 
is to reduce a planetary absorption signature when measuring 
within a stellar line and comparing to a nearby continuum, (3) 
the effect is small for a typical stellar absorption line (\ensuremath{\sim} 
-0.004\%), and (4) the strength of the effect is consistent with 
theoretical 1-D stellar atmosphere predictions. Although the 
1-D stellar model used in our analysis does not contain a stellar 
chromosphere and assumes local thermodynamic equilibrium, the 
similarity between the model and data suggests that their effects 
are negligible on the transit light curve at these wavelengths 
and resolutions.

\subsection{Performance of solar-like Atlas limb-darkening models}

To gauge the performance and limitations of Atlas atmospheric 
models, in relation to analyzing planetary transits, we compared 
the limb-darkening predictions of a solar Atlas model to the 
measured values of the sun. The sun represents 
the only test case where sufficiently high-precision optical 
limb-darkening data exists.  HD209458 is a solar-like star, being 
classified as a G0 V, while the sun has a similar spectral type and is a G2 V.  In 
analyzing the performance of model atmospheres, Bertone et al. 
(2004) found Atlas models to provide a good fitting accuracy 
for physical parameters, namely T$_{eff}$ and log g, for early type 
main sequence stars, B to F, down to late G/early K stars. For 
late K and M stars, the models produce systematically poorer 
fits, due largely to the incomplete treatment of molecular opacity, 
which becomes important in cooler stars. As HD209458 is slightly 
warmer than the sun, molecular opacity should correspondingly 
play a slightly weaker role, and the performance of the Atlas 
model itself should be similar for such main-sequence stars. 
Testing a solar Atlas model vs. solar data represents a test 
of the model itself and the assumptions within the model, as 
the physical parameters of the sun, which are input into the 
atmospheric model, are both accurately and precisely measured. 
The uncertainty in the measured stellar parameters themselves, 
and its effect on limb-darkened corrected transit light curves, 
is a separate issue dealt with in \S 4.1. 

Seen in Fig. 5 are high precision solar limb-darkening measurements 
for 25 precisely chosen continuum wavelengths between 3,000 and 
11,000 {\AA} from Neckel \& Labs (1994; hereafter NL94) and the 
corresponding solar Atlas predictions, taken from the non-linear 
solar Atlas monochromatic coefficients computed by Claret (2000). 
The Atlas model was calculated at a low resolution, so those 
wavelengths which matched closest were chosen. The Atlas model 
is seen to perform well over \textit{\ensuremath{\mu}} values between \ensuremath{\sim}0.2 
and 1.0, slightly overestimating the strength of limb-darkening, 
consistently at every wavelength, by a couple of percent (see 
Fig. 6). Our Atlas-solar comparison agrees with earlier 
studies (Castelli, Gratton \& Kurucz 1997). The general overestimation 
by atlas models is expected, due to the inherent limitations 
of 1-D models when describing purely 3D turbulent effects. This 
overestimation has been observed in both a F5 V and K1 V star 
when comparing Atlas models and 3-dimensional hydrodynamical 
models to high-precision interferometric measurements (Aufdenberg 
2005; Bigot et al. 2006). The largest differences between the 
model and data exist at the very limb, where Atlas models predict 
a dramatic increase in limb-darkening strength, a trend not observed 
in the solar limb-darkening data.

\subsection{Simulated Solar Transits}

The effect of adopting an Atlas 
model, to describe limb-darkening in transit light curves, can 
be seen by simulating transits with the high precision NL94
solar data and comparing them to transits using an Atlas solar 
model. For this purpose, we modeled transit light curves using 
the best-fit parameters of the HD209458 system, and both the 
nonlinear limb-darkening coefficients predicted by the Atlas 
solar model as well as the solar data (see Fig. 7). With 
overall stronger limb-darkening, the Atlas models produced systematically 
deeper transits between 2$^{nd}$ and 3$^{rd}$ contact and shallower light 
curves during ingress and egress. The largest differences between 
the solar and Atlas transits occur during ingress and egress 
where transit parameter fits are relatively insensitive to the 
planet-to-star radius contrast, $R_{pl}$/$R_{star}$, which 
is the desired parameter for this study. This insensitivity combined 
with its small integrated flux contribution of the limb, in contrast 
to the rest of the stellar surface, limits the overall affect 
of the observed large limb model deficiencies during mid-transit. 
At mid-transit, a ${\sim}$2\% overestimation in limb-darkening 
(corresponding to $\mu\sim$0.5) translates to 
an extra transit flux drop of ${\sim}$0.01\%. These trends are 
seen to be largely reproducible, wavelength-to-wavelength, at 
a ${\pm}$0.005-0.006\% level. The NUV, where the strength of 
limb-darkening is larger, also exhibits this general overestimation 
trend, but with a slightly larger scatter compared to the rest 
of the optical.

We corrected the transit light curve, computed with the NL94 
data, for limb-darkening effects as predicted by the Atlas model. 
This procedure is comparable to our limb-darkening correction 
procedure adopted for HD209458b, i.e. using the predicted limb-darkening 
from Atlas models to correct transit light curves which are the result
of the actual limb-darkening of a star. These corrected light 
curves can be seen in the top curves of Fig. 8, where the 
limb-darkening overestimation by the Atlas models translates 
into a slightly lower flux ratio at mid-transit, and slightly 
higher flux ratio just before 3$^{rd}$ contact, or just after 2$^{nd}$ 
contact, with this general trend seen at all wavelengths. Thus, 
introducing the Atlas model produces two systematic effects ($A$)
a wavelength-to-wavelength flux scatter of up to ${\pm}$0.005-0.006\% 
and ($B$) a phase-dependent trend of overcorrecting the transit 
flux during the middle ${\sim}$2/3 of in-transit times and under-correcting 
during the remaining ${\sim}$1/3. The first systematic effect 
is likely due to the internal accuracy of the Atlas models, where 
uncertainties in opacity database tables and metal line-blanketing, 
for instance, could reveal themselves. The second systematic 
effect is clearly related to the general trend of Atlas models 
over predicting the strength of limb-darkening, at a similar 
level, in all wavelengths and tied to the treatment of convection 
in one-dimensional atmospheric models.

To test if the broadband NUV and Na absorption features seen 
in our data could successfully be recovered using Atlas model 
limb-darkening corrections, we also simulated transits using 
the NL94 data, which contained an extra 0.01 R$_{JUP}$ in planetary 
radii for the NUV and an extra 0.02 R$_{JUP}$ in planetary radii for 
selected Na-like wavelengths. These simulated absorption transit 
light-curves were then corrected with those generated with solar 
Atlas models and the best-fit planetary radius (see Fig. 8). The 
two simulated absorptions can easily be identified above 
that of the zero absorption case, with the general phase-dependent 
trend largely preserved. Therefore, when averaging over a series 
of in-transit spectra and comparing absorption values at different 
wavelengths, the values of the {\it relative} absorption differences 
will largely be retained after the Atlas corrections, but the 
flux level itself could change slightly, depending on the phase 
sampling (effect $B$). At the extremes, using only data 
at mid-transit, 2$^{nd}$ contact, or 3$^{rd}$ contact, this absolute flux 
difference is seen to be up to \ensuremath{\sim}0.01\%. These trends 
are observed in our absorption simulations. When we average over 
the in-transit phases, and recover the input absorption signal 
(see Fig. 9), we get relative absorption values (above the 
Atlas corrected zero absorption case) of 0.026 ${\pm}$0.003\% 
and 0.046${\pm}$0.002\% for the NUV-like and Na-like signature 
respectively. These values closely match the input 0.022\% and 
0.044\% values, indicating that Atlas model corrections can retain 
relative absorption signals at level of around \ensuremath{\pm}0.003\%. 
Selecting subsets which contained only similar orbital phases 
produced equivalent differential results, indicating effect $B$ 
has a limited impact on relative absorption measurements.

\subsection{Application of solar test to HD209458}

The residual G430L and G750L Atlas-fit light curves, additionally see Fig. 
4 of Knutson et al. (2007), show the same trends as observed 
in our solar test. Namely, the HD209458 transit light curves 
show a systematically slightly higher residual flux at mid-transit 
compared to 2$^{nd}$ and 3$^{rd}$ contact, at a \ensuremath{\sim}0.01-0.02\% flux 
level as in our solar test, with this trend repeated at all optical 
wavelengths. This indicates the performance of the Atlas models 
is similar for both HD209458 and the sun. When fitting for the 
planetary radius, in the case of HD209458, adopting an Atlas 
limb-darkening model would therefore seem to systematically fit a slightly 
smaller $R_{pl}$/$R_{star}$ ratio at every wavelength, and thus 
result in smaller planetary radii (given uniform and complete 
phase coverage). However, this systematic error is much smaller 
than the uncertainty due to the stellar mass and has a negligible 
effect on the final determined planetary radius. This is reflected 
in the analysis of Knutson et al. (2007), who find planetary 
radii values which differ by less than 1$\sigma$, when using either the 
method of adopting Atlas models or fitting for limb-darkening 
coefficients.

Our analysis shows that, over the entire optical regime 3,000-10,000 
{\AA}, it is possible to use Atlas models to accurately measure 
the {\it relative} value of planetary radii to around a 0.005-0.006\% 
precision level in solar-type stars. This precision level is 
comparable to our highest S/N achieved when averaging over large 
wavelength ranges (\texttt{>}1000 {\AA}) and multiple exposures (see 
\S 4.2). Thus, the systematic errors introduced 
by assuming Atlas models are negligible for our purposes of probing 
relative transit depths for signals \texttt{>}0.01\%. The success 
of using Atlas models in analyzing transits of the K0 V star 
HD189733 (Pont et al. 2007), further highlights their capabilities. 
As a relatively cooler K star, molecular opacity plays a correspondingly 
larger role in HD189733 than the sun or HD209458, yet an Atlas 
model provides superior fits compared to leaving the limb-darkening 
coefficients as free parameters in transit fits (F. Pont, private 
communication). As more high-precision measurements become available 
across a wider wavelength range, care will have to be taken when 
comparing transit parameters determined by separate analysis, 
which make different limb-darkening assumptions.

\section{Analysis}
\subsection{Stellar parameter uncertainties}
Uncertainty in the stellar parameters (T$_{eff}$, log g, log$_{10}$[M/H], v$_{turb}$) can lead to a range 
of possible stellar models and thus a range of predicted limb-darkening. 
As limb-darkening derives mainly from intensity differences between 
blackbodies at different temperatures, changes in the T$_{eff}$ of 
the star have the largest effect on the calculated limb-darkening. 
For HD209458, the range of T$_{eff}$ found in the literature ranges 
from \ensuremath{\sim}6,000-6,100 K (6,030\ensuremath{\pm}140 K Allende Prieto 
\& Lambert 1999; 6,000\ensuremath{\pm}50 K Mazeh et al. 2000; 6,117\ensuremath{\pm}26 
K Santos, Israelian, \& Mayor 2004; 6,099 K Fischer \& Valenti 
2005) and we have adopted a T$_{eff}$ of 6,100 for this study.

We estimate the uncertainty of limb-darkening inferred from 
stellar parameter uncertainty by estimating how large a temperature 
increase or decrease is necessary to produce a difference flux of 0.02\% 
at mid-transit between bandpasses, which is near the 
absorption signal detected in the NUV region. In our wavelength 
regime, the flux at mid-transit in the NUV is the most sensitive 
to changes in limb-darkening, making this test an upper-limit 
to the T$_{eff}$ change needed at longer wavelength bandpasses and 
other phases. From Claret (2000), we use non-linear limb-darkening 
for the photometric bandpasses U and B calculated at three different 
temperatures, and measure the transit depth that results. The 
U and B wavelength ranges roughly correspond to the \ensuremath{\sim}0.03\% 
flux difference observed in our transmission spectrum between 
the NUV (3,000-3,700 {\AA}) and nearby optical (4,000-5,500 {\AA}). Using 
the best-fit HD209458 parameters, we calculate that a T$_{eff}$ change 
from 6,000 K to 5500 K changes the U band mid-transit depth by 
0.05\% and the B band by 0.03\%, resulting in a \ensuremath{\sim}0.02\% 
change in relative flux depth. Likewise, changing the T$_{eff}$ from 
6,000 K to 6,500 K changes the mid-transit depth by 0.05\% in the 
U band and 0.03\% in the B band, also resulting in a \ensuremath{\sim}0.02\% 
change. Thus, an uncertainty of only \ensuremath{\pm}50 K for the T$_{eff}$ 
of HD209458 corresponds to a NUV limb-darkening uncertainty of 
only \ensuremath{\sim}0.002\%, a level well below our S/N level.

\subsection{Red noise}

We quantify the level of systematic `red noise' in our transmission 
spectrum following the procedures detailed in Pont, Zucker, \& 
Queloz (2006). In the absence of red noise, averaging over successive 
measurements increases the S/N in a predictable manner, with 
the standard deviation of the average varying inversely with 
the square root of the number of measurements, N$_{t}$. In 
the presence of red noise, however, the standard deviation does 
not decrease as quickly, when averaging over larger number of 
measurements, and in the limit of completely correlated noise, 
becomes a constant of N$_{t}$.

To estimate the level of correlated noise in our data, we take 
the out-of-transit spectra, binned over N$_{w}$ pixels in wavelength, 
and calculate the variance and standard deviation from a distribution 
of points binned by N$_{t}$ successive measurements. This is 
repeated for an increasingly larger number of N$_{t}$ and N$_{w}$
bin sizes. We checked different binning combinations to ensure 
correlated noise did not appear at our binning frequency. Plotted 
in Fig. 10 is a sample of red noise estimation from the 
G750L data in the 6,200 {\AA} region. Typically, photon noise is 
seen to dominate down to precision levels of \ensuremath{\sim}0.02\%, 
where the effects of red noise start to become visible. For most 
of the optical, binning over large wavelength regions and multiple 
exposures allows us to reach precision levels of \ensuremath{\sim}0.006\%, 
although after binning beyond \ensuremath{\sim}30 measurements, the S/N 
ratio usually does not significantly increase.

As binning over larger wavelength bands has diminishing returns,
the presence of red noise gives our transmission
spectra a similar S/N level at moderate wavelength bins as compared to
those larger wavelength bins.  
This is reflected in Fig. 11, where all five 1$\sigma$ error bars calculated to show the significance of
observed broadband features are similar to each other, though they differ in wavelength breadth. 
The broadband errors are also similar to the typical 1$\sigma$ uncertainty levels when binning over 16 pixels, 0.011\%.

\subsection{Broadband absorption signatures, optical transit transmission spectra}

Using the G750M and G750L spectra, we used a limb-darkened corrected 
transit spectral ratio to probe the full wavelength dependence 
of atmospheric absorption features, producing an optical transmission 
spectra. The relative planetary radii, when comparing between 
different wavelengths, are largely independent of the uncertainties 
of the stellar parameters and can be determined much more precisely 
than the actual radii values themselves (Knutson et al. 2007). 
Putting the low resolution G750L data together with the G430L 
data (Ballester, Sing \& Herbert 2007), allows a probe of the 
region from 3,000-10,000 {\AA} with the Na D lines located in a 
high S/N region in the middle. 

Our low resolution transmission spectra was produced by interpolating 
all the systematic error corrected spectra onto a common wavelength 
scale. We then produced both an average out-of-transit spectra 
and an average limb-darkened corrected ``in''-transit spectra 
(using phases between 2$^{nd}$ and 3$^{rd}$ contact), with the ratio 
of the two providing our ratio spectrum. The error was computed 
taking into account both photon noise and red noise, as outlined 
in section 1.7. The G750L spectral region blue-ward of 5,548 
{\AA} at the edge of the spectrum was omitted as it contains a 
region of very rapidly decreasing response whose spectral ratio 
value is 0.03\% lower, when compared to the higher S/N overlapping 
G430L region. This omitted G750L region is seen to have a larger 
amount of red noise, limiting the accuracy levels in the region 
to \ensuremath{\sim}0.02-0.03\%, which likely explains the flux difference 
between the two data sets. We used the best fit value of 1-($R_{pl}$/$R_{star}$) 
\ensuremath{\sim} 1.455\% from \cite{Knut} as our baseline transit 
depth reference, subtracting our spectral ratio from that value 
to construct a plot of the atmospheric absorption profile (Fig. 
11). We also performed the same procedure on the G750M data, 
though a shift of +0.00023 was needed to generate an average 
absorption level consistent with that of the low resolution data. 
As the medium resolution data was taken three years before the 
low resolution data, long-term differences in star spots, for 
instance, could easily account for the small measured difference 
in absolute absorption level. 
The effects from the uncertainties in the transit parameters 
were tested by varying the period, inclination, stellar radius, 
stellar mass by 1$\sigma$ (as reported in Knutson et al. 2007) and producing 
a new transmission spectra. Due to limb-darkening, changing the 
inclination is seen to have a larger affect than the other parameters, 
though all produce negligible affects on the final transmission 
spectrum. For instance, even for the NUV at 3,000 {\AA} which contains 
strong limb-darkening, changing the inclination by 1$\sigma$ still only 
has a 0.002\% effect on the final transmission spectrum.

The wavelength range of the G750L grating extends in the red past 1$\mu$m, though
with significant detector fringing \citep{Knut}.  The data reduction procedures
used here were insufficient to correct for the large fringing effect on a pixel-by-pixel
basis.  Consequently, those regions which showed signs of significant fringing were 
left out of this analysis, corresponding to wavelength regions longer than $\sim$8,000{\AA}.  

\section{Discussion and Conclusions}

Our optical transit transmission spectrum combining all three gratings (see Fig. 11), gives
a consistent and more complete view of the observed absorption features present in HD209458b. 
The low resolution transmission 
spectrum reveals three prominent broadband absorption features:
(1) a NUV absorption below ${\sim}$4,000 {\AA};
(2) a strong broadband Na feature;
and (3) an additional absorption feature at ${\sim}$6,250 {\AA}.
The medium resolution data shows: (4) the strong Na core absorption ($\sim$0.065\% in a 4.4{\AA} line-core band) and (5) confirms the presence
of the absorption feature at ${\sim}$6,250 {\AA}. 
The NUV absorption region was first reported by \cite{Ballester07}, and is interpreted along with wavelengths shorter than
5,000 {\AA} by \cite{Lecavelier08} to be Rayleigh scattering by H$_2$.  While the long wavelength continuum redward of $\sim$6,200 {\AA} can be
explained by absorption by TiO and VO \citep{desert08}.

Seen at low resolution, the wide Na wings and additional absorption at 6,250{\AA} place the medium resolution
data on an apparent plateau, ${\sim}$0.045\% above that of the minimum absorption levels seen around 5,000{\AA} and redward of 7,000{\AA}.  
As such, the total Na core absorption reaches to $\sim$0.11\% within a 4.4{\AA} band above the minimum level at 5,000{\AA}.
This strong Na absorption is much closer to the original cloudless predictions \citep{SS00,Brown01,Hubbard01}, albeit with a
more complicated Na line absorption profile \citep{Sing08}.
The large Na line-core absorption would seem to disfavor high altitude clouds or haze cutting into the 
absorption signature, as recently seen on HD189733 by \cite{Pont07}.    
The sharp transition from broad Na features at lower absorption depths to sharp narrow features at high absorption depths, however, indicates
condensation of Na is likely in the upper atmosphere, which would deplete those higher altitudes of atomic Na \citep{Sing08}.

The complicated Na~D line shape and telluric contamination illustrate some of
the difficulties in observing this atmospheric absorption signature in
HD209458b.  Ground based spectroscopic measurements, such as the Na detected in HD189733b \citep{Redfield}, require
airmass corrections and are only sensitive to the Na~D line
cores where the atmospheric atomic Na in HD209458b is substantially depleted.
Furthermore, telluric Na contamination has a non-negligible impact on
the signature, further masking the small signal. 

As seen here with Na, transit spectra at multiple resolutions are needed to properly interpret complex
transmission features.  Data at low resolution, covering a broad wavelength range, places
the overall features in context, while data at medium or high resolution is better suited
to identify specific absorption features and probe planetary atmospheres over a large pressure
ranges.  Our tests show that for exoplanets with solar type stellar hosts, such transmission spectra can be 
corrected for limb-darkening effects to a high precision with current atmospheric models.

Space-based observations on exoplanetary transits offer the extraordinary ability
to provide exoplanetary transmission spectra at multiple resolutions,  
leading to detailed atmospheric composition information.  As the signal-to-noise, resolution,
and wavelength coverage of such spectra improves, multiple species can be
detected and atmospheric properties inferred.  The data quality of the spectra used
here and results herein help indicate the full potential of the transit method, which can
be further exploited with the impending repair of HST and future launch of the
James Webb Space Telescope.  

\acknowledgments
D.K.S. is supported by CNES.  This work is based on observations
with the NASA/ESA Hubble Space Telescope, obtained at the
Space Telescope Science Institute (STScI) operated by AURA, Inc.  
Support for this work was provided by NASA through a grant from the STScI.
We thank F. Pont, D. Ehrenreich, G. H\'{e}brard, R. Ferlet, and F. Bouchy for discussions and insight.
We also warmly thank our referee for their comments and corrections.


\begin{figure}
  \plotone{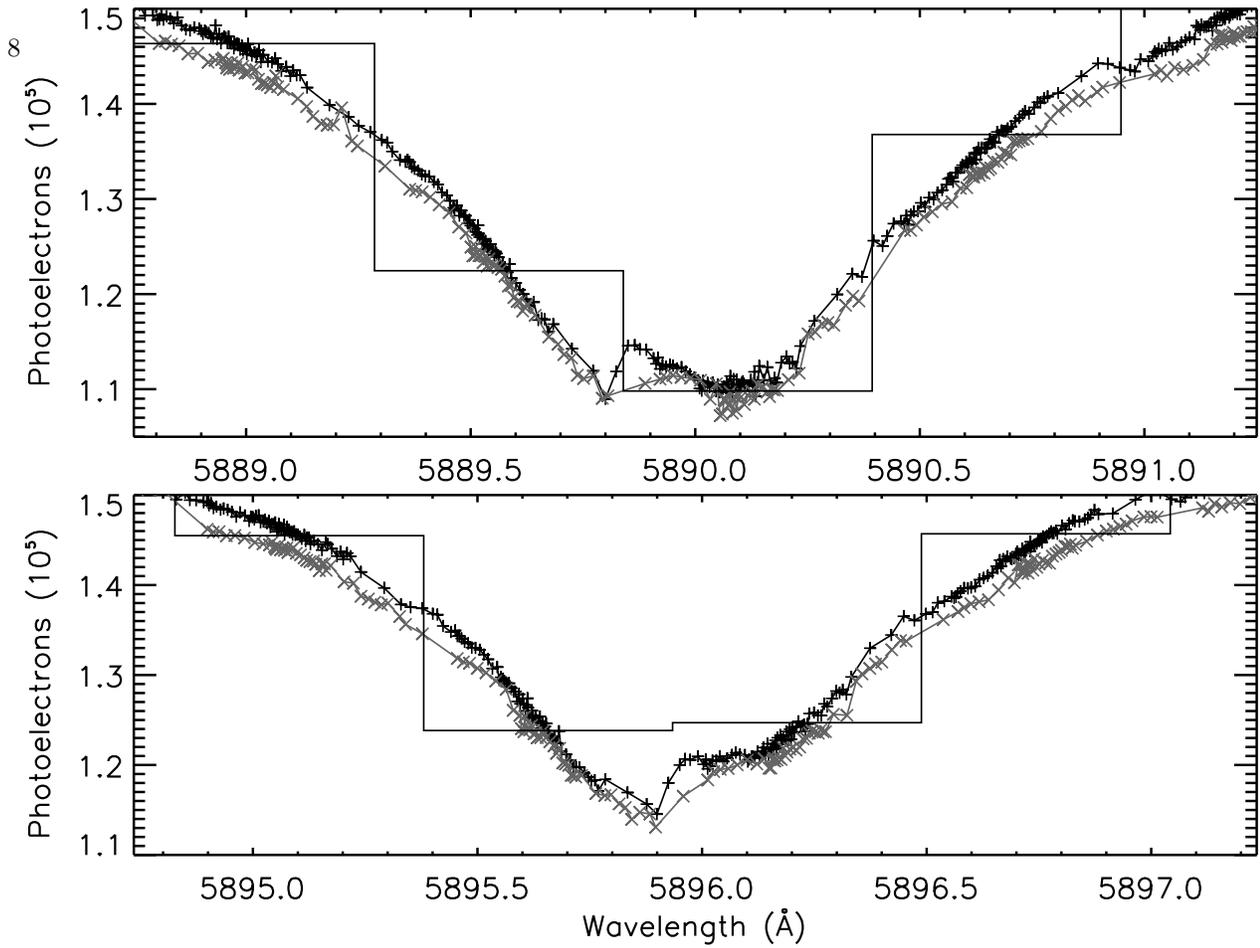}
  \caption{ G750M spectra of  the Na D2 (top) and D1 (bottom) line cores. Plotted
  are all  of the out-of-transit  spectra (black crosses), as  well as
  the  in-transit  spectra  (grey X).  The uncertainty in counts for each point is dominated
  by photon noise ($\sim\pm$400 photoelectrons) and in is smaller than the plotted symbol size.
  The first out-of-transit
  spectra in the series is plotted  in histogram mode to show the size
  of  each  pixel,  0.554 {\AA}.  Plotted  together,  each  individual
  spectrum's Doppler and pixel-shifts  (accurate to $\sim$0.1 pixel) effectively add up to better sample the
  PSF of the stellar absorption line profile.  The
  $\sim$1.5\% difference  in flux  during transit can  be seen  in the
  flux difference  between the two spectra.  However, the regions showing contamination from 
  5889.5 to 5889.7 {\AA} and from 5,895.5 to 5,895.8 {\AA} shows  little-to-no difference in  flux
  during transit.  This region shows variations with the  HST orbital
  period,   which we attribute to terrestrial-based sodium
  contamination.}
\end{figure}
\begin{figure}
  \plotone{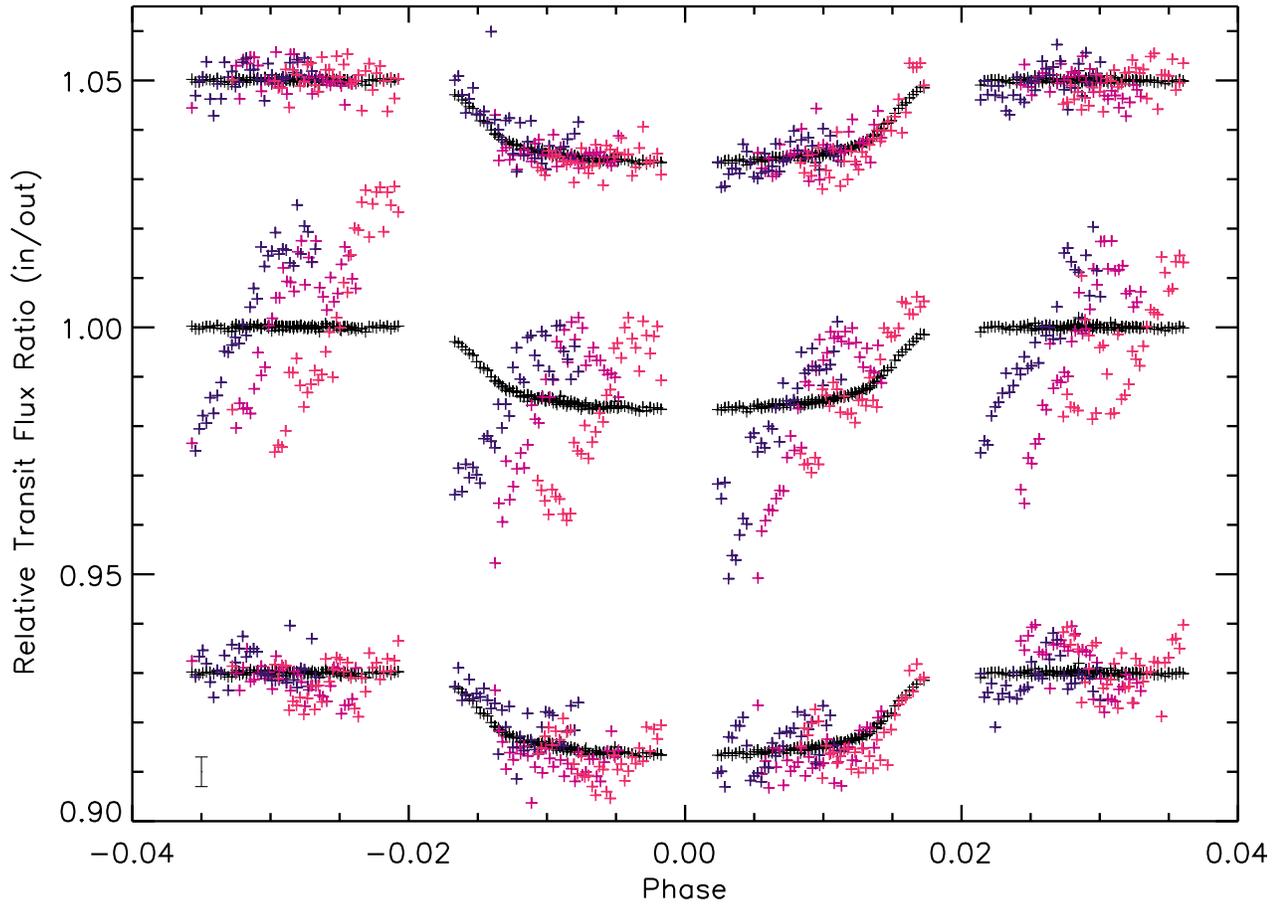}
  \caption{G750M   photometry   of   the   contaminated
  region.  Plotted are photometric curves from the comparison
region  which  includes  wavelengths between the Na lines 5,818-5,843  {\AA}  and
5,943-5,968 {\AA}  (black) along with (top) one pixel between the Na lines at 5,893.4
{\AA} which is completely uncontaminated.  The plot also
shows (middle) the same comparison region along with a highly
contaminated pixel in the Na D1 line core at 5,895.65 {\AA} and (bottom) 
a nearby D1 line core pixel which is largely uncontaminated.  
A representative error bar for the single pixel photometric light curves is also shown in the lower left-hand
corner, the error for the comparison region is significantly smaller than the plot symbol size.
The three colors (purple, blue, cyan) distinguish the 3 different HST visits,  each
consisting of 5 consecutive orbits (4 orbits plotted).  The telluric
contaminated Na core can be seen as a \ensuremath{\sim}4-5\% change in
flux which modulates on the HST orbital cycle, repeating each orbit.}
\vspace{0.3cm}
\end{figure}

\begin{figure}
  \plotone{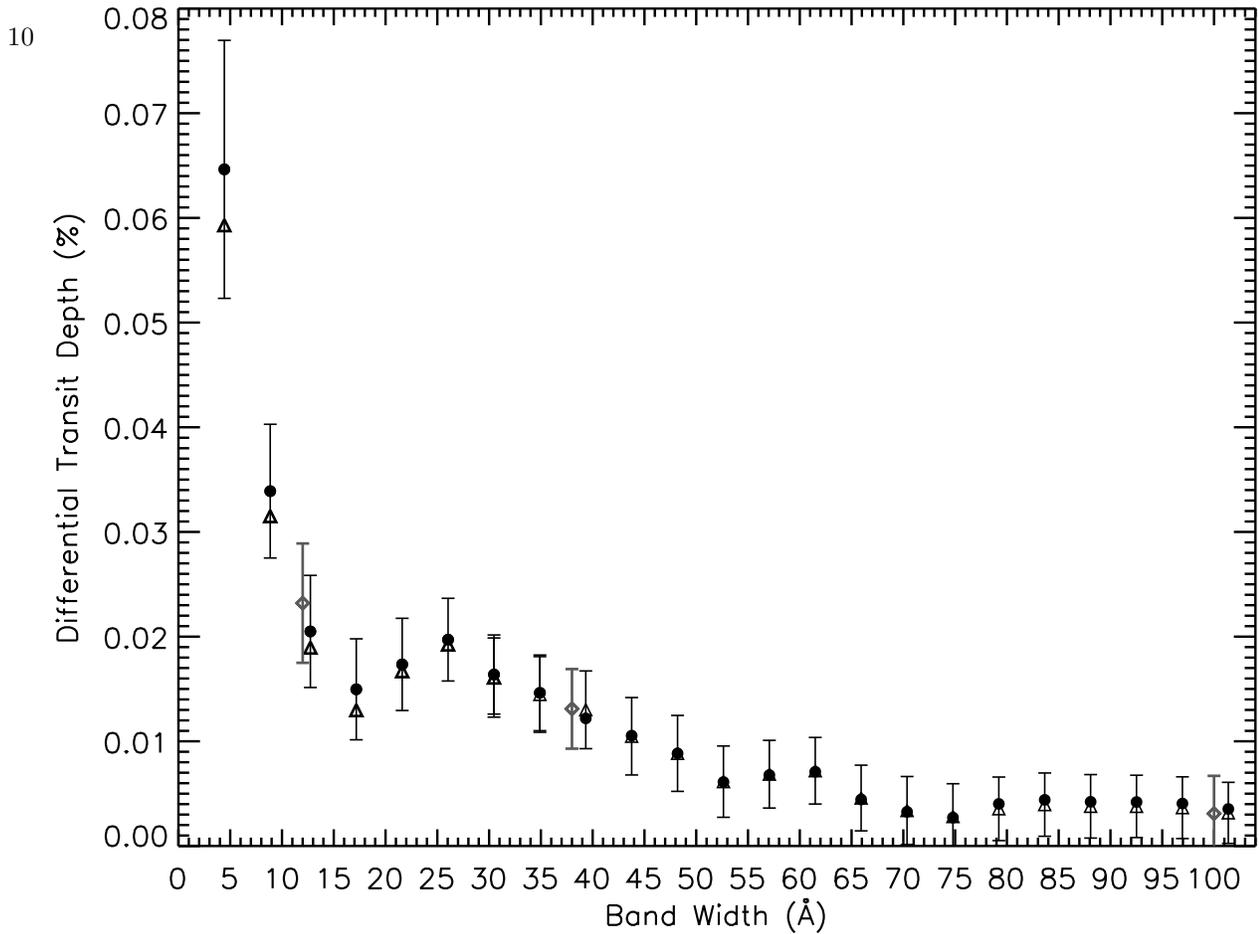}
  \caption{Double  differential  photometric   Na  core
  absorption.  Plotted are  the measured  absorption of  various size
wavelength bands, each centered on the Na D lines, compared to that of
a  ``wide''   comparison  region  composed  of   the  regions  between
5,818-5,843 {\AA} and 5,943-5,968 {\AA} (triangles). The limb-darkened
corrected points (circles), with the  associated 1$\sigma$ error bars, and the
original three  Charbonneau et al.  (2002) measurements (grey-diamonds)
are also plotted.  For each of our 23  wavelength bands, we integrated
the spectral  flux to produce transit light  curves. These photometric
light curves were then subtracted from the light curve of the ``wide''
comparison  band.  The atmospheric  sodium absorption,  and associated
error, from  each band  were then measured  by mean of  the difference
light  curve from  the  in-transit phases,  between  second and  third
contact.   The   two  narrowest  bands  are   corrected  for  telluric
contamination and reveal a stronger Na D core absorption.}
\end{figure}

\begin{figure}
  \plotone{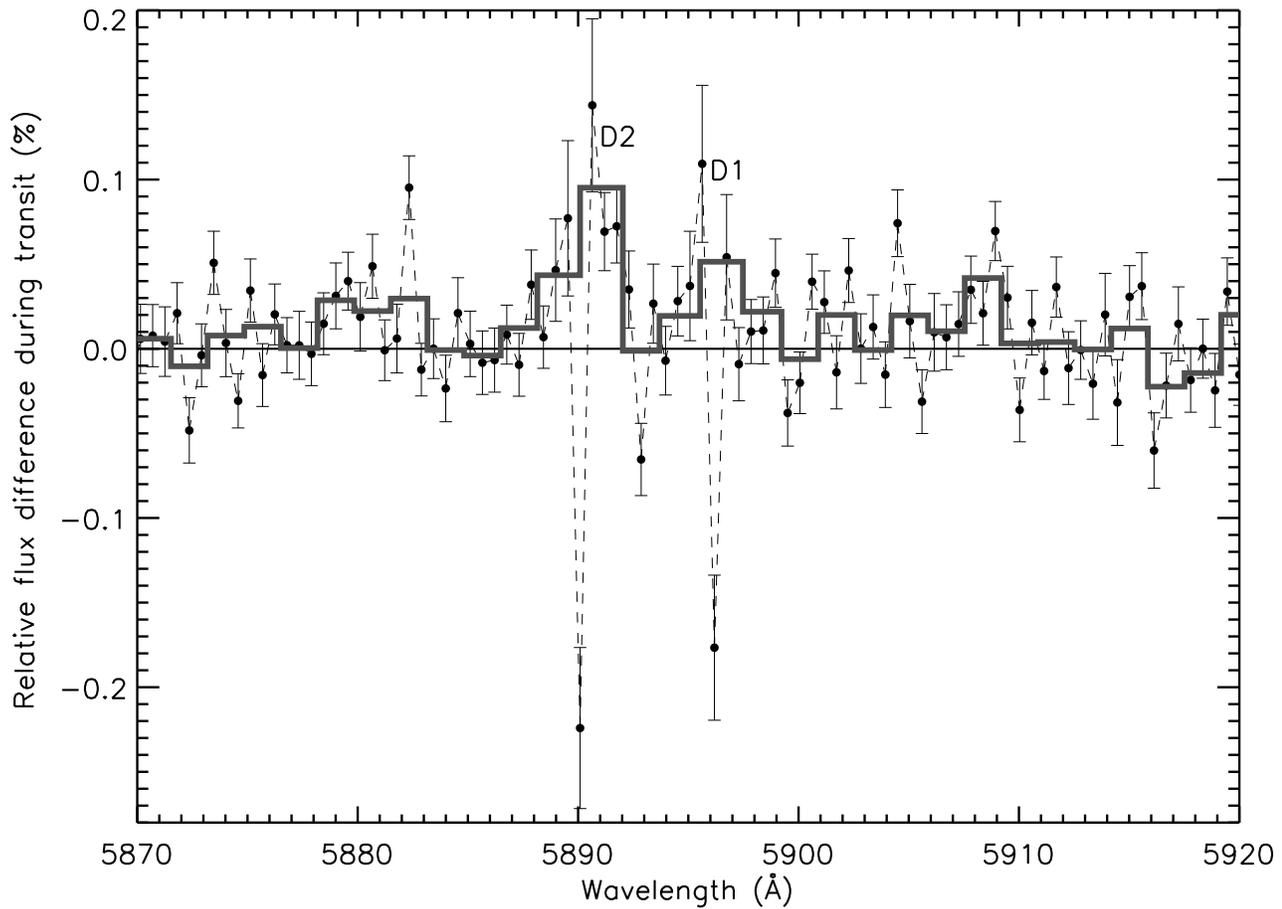}
  \caption{Double     differential    spectroscopic
  measurement.  Plotted   is  the  differential   absorption  at  each
  wavelength in the  Na region compared to the ``wide'' comparison band
  along  with  the   associated  1$\sigma$  error.  Over-plotted  (grey
  histogram)  is the  data binned  over 3  pixels with  the  two large
  negative points, contaminated by terrestrial absorption, excluded in
  the bin. The Na D1 and D2 doublet is resolved and seen as the two largest
  absorption peaks.}
\end{figure}
\begin{figure}
  \plotone{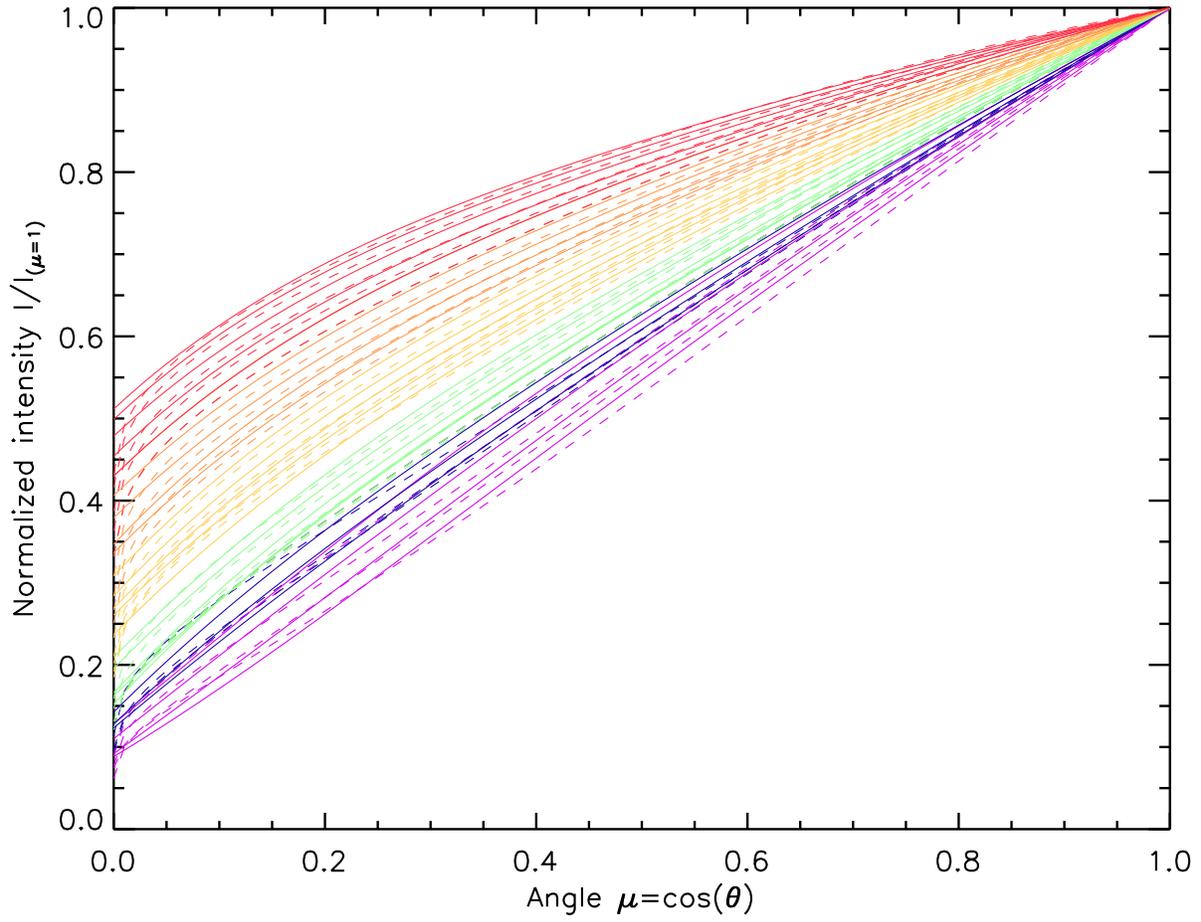}
  \figcaption{Solar  limb-darkening data  and  Atlas model.
Plotted are  23 optical solar limb-darkening  measurements from Neckel
\& Labs (1994;  solid lines) along with those of  an Atlas solar model
(dashed lines). The wavelengths of each curve range from 3204.68 {\AA}
(bottom purple curve) to 10989.5 {\AA} (top red curve).}
\vspace{0.4cm}
\end{figure}

\begin{figure}
  \plotone{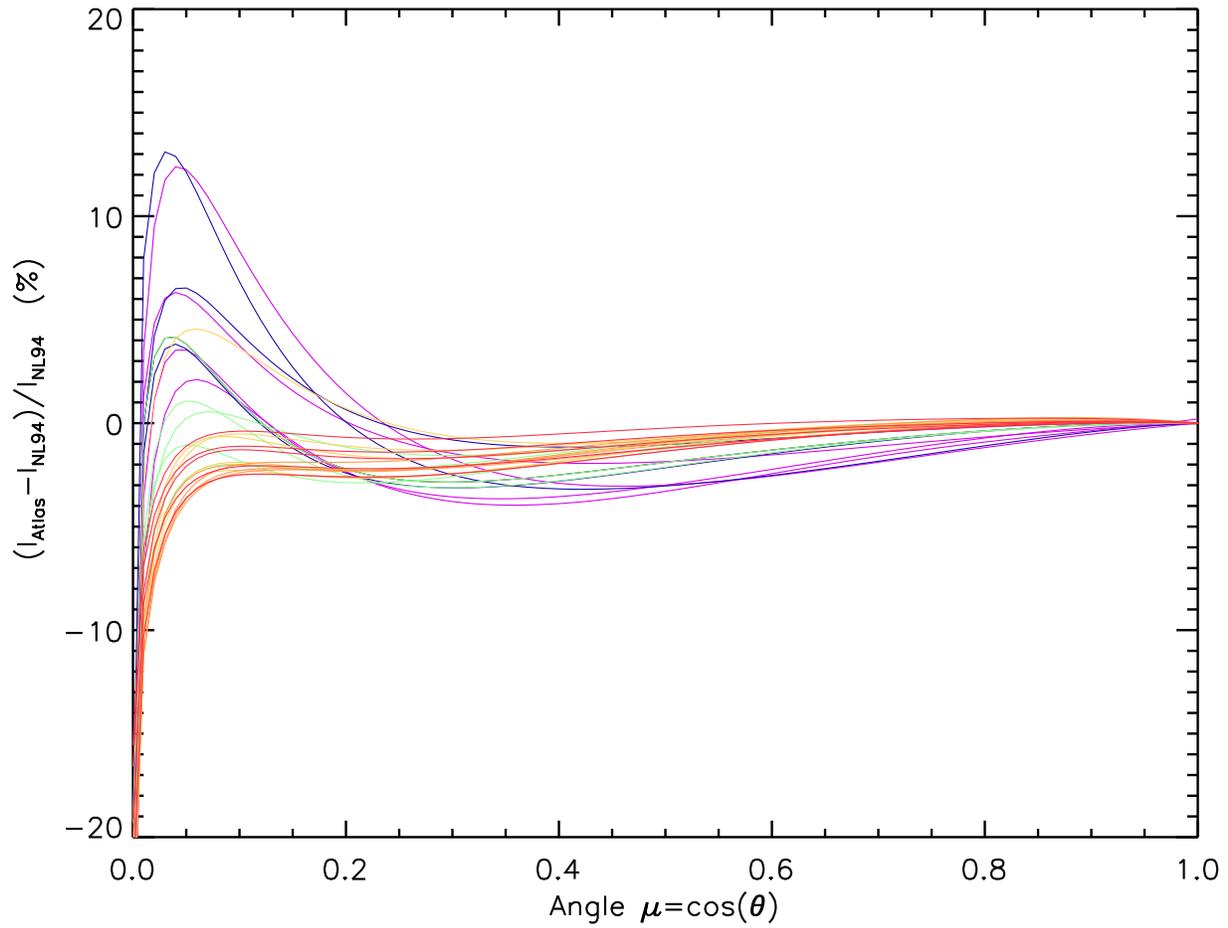}
  \figcaption{Difference  between solar  limb-darkening data
  and   Atlas  model.    Plotted   is  the   difference  between   the
  limb-darkening curves  predicted by the Atlas model  and measured by
  NL94.  For $\mu$ values between  $\sim$0.2 and 1.0, the model tends
  to over  estimate the strength  of limb-darkening by a  few percent.
  The greatest difference is observed to be at the very limb, were the
  models   are  known   to   have  their   greatest  uncertainty.}
\end{figure}

\begin{figure}
  \plotone{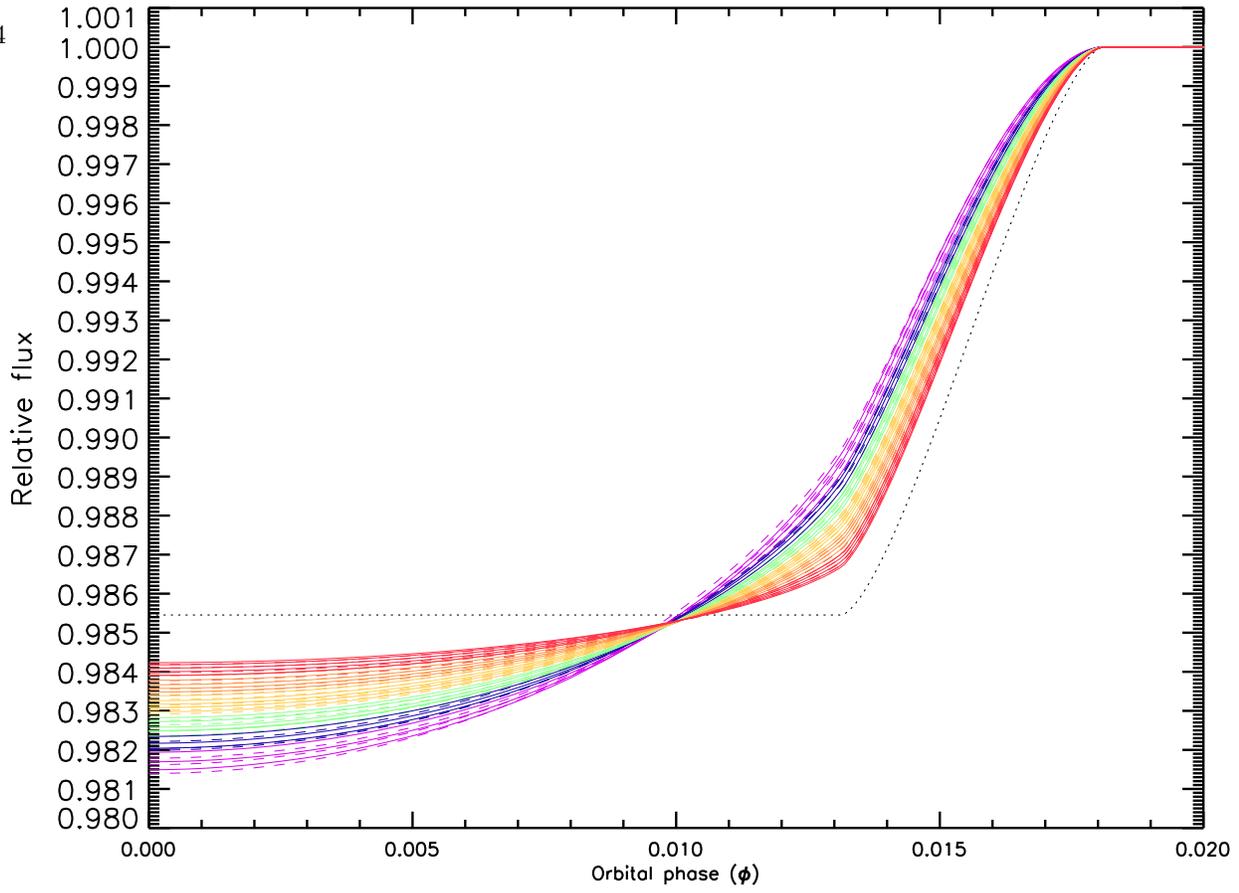} 
  \figcaption{Simulated solar  transits.   Plotted are  the
  simulated  transits   using  the  best-fit   transit  parameters  of
  HD209458,  along  with the  solar  limb-darkening  measured by  NL94
  (solid lines)  and a  solar Atlas model  (dashed lines).  The dotted
  line  represents  the  theoretical  transit  without  limb-darkening
  effects.}
\vspace{0.4cm}
\end{figure}

\begin{figure}
  \plotone{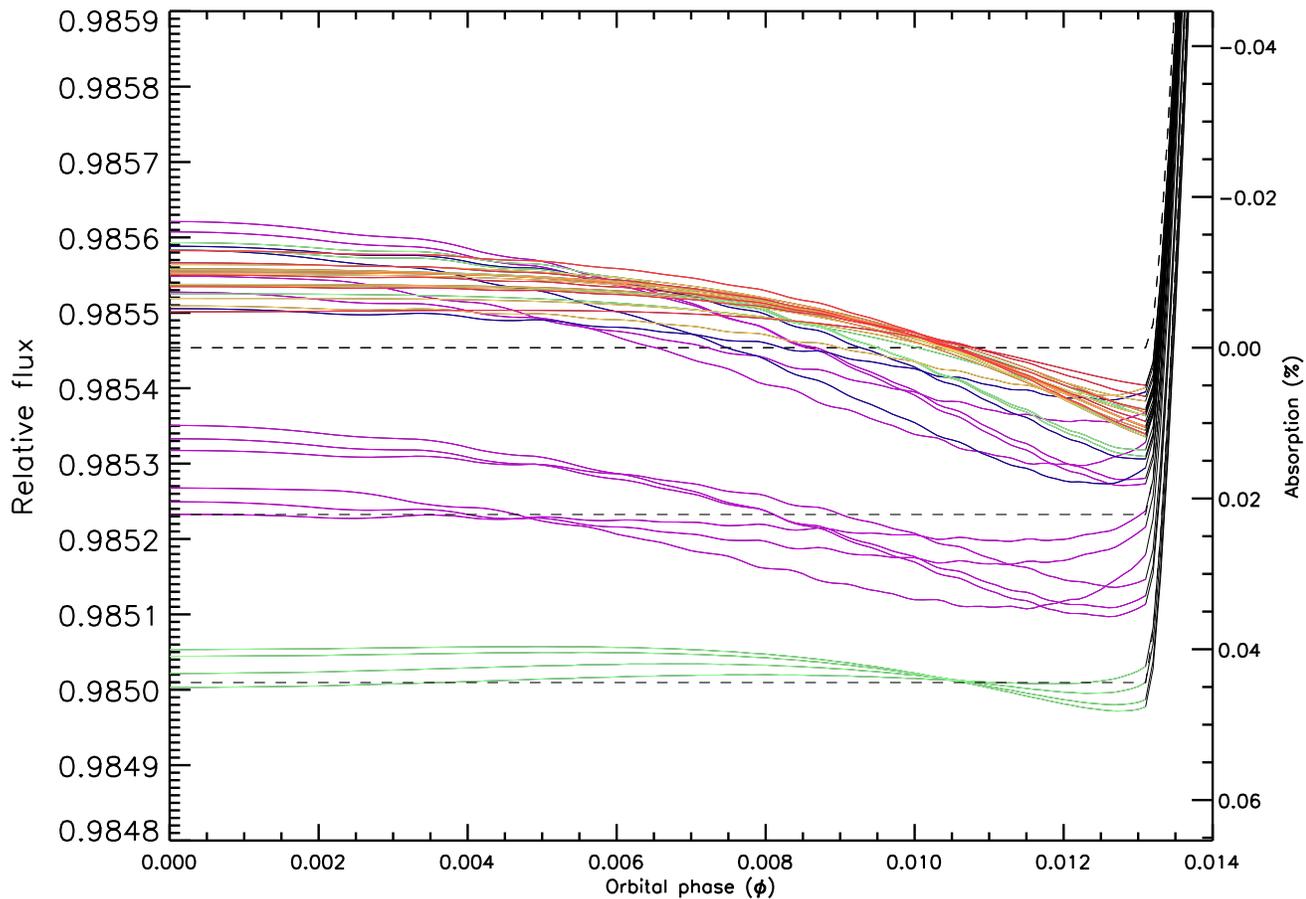}
  \figcaption{Simulated   solar  transits
  corrected  by  Atlas   limb-darkening.   Plotted  are  the  transits
  generated with NL94 limb-darkening data and corrected with a solar
  Atlas  model.   Those  simulated  transits containing  the  best-fit
  planetary radius (1.32 R$_{Jup}$; corresponds to 0\% absorption) are
  plotted (top).  Also plotted, are simulated transits with a NUV-like
  absorption signature,  (middle, purple; calculated  with a planetary
  radius of 1.33 R$_{Jup}$), and simulated Na-like signature, (bottom,
  green;  calculated  with  a   1.34  R$_{Jup}$).   The  dotted  lines
  represent  the  theoretical transits  for  all  three  radii in  the
  absence of limb-darkening effects.}
\end{figure}

\begin{figure}
  \plotone{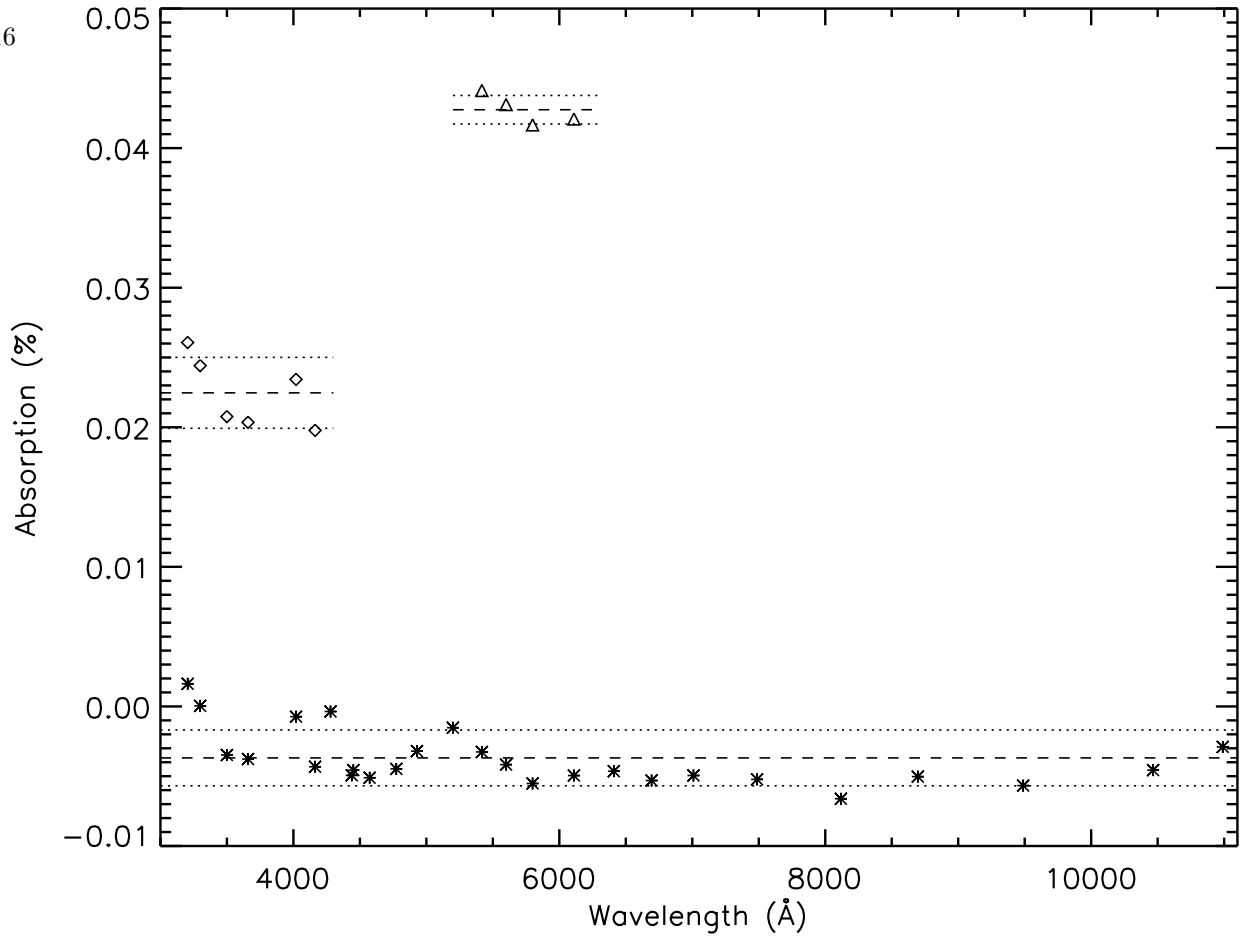}
  \figcaption{Simulated solar  transit  ratio spectra  with
  NUV-like and Na-like signatures.  Plotted are the average in-transit
  values from  the Atlas corrected  simulated solar transits  (Fig. 8),
  which  originally  contain   zero  absorption  (stars),  a  0.0022\%
  NUV-like  signature  (diamonds), and  a  0.0044\% Na-like  signature
  (triangles).  The  two simulated differential  absorption signatures
  are easily  recovered.  For  each set, the  average and  1$\sigma$ standard
  deviation are also plotted (dashed and dotted lines).}
\end{figure}
\begin{figure}
  \plotone{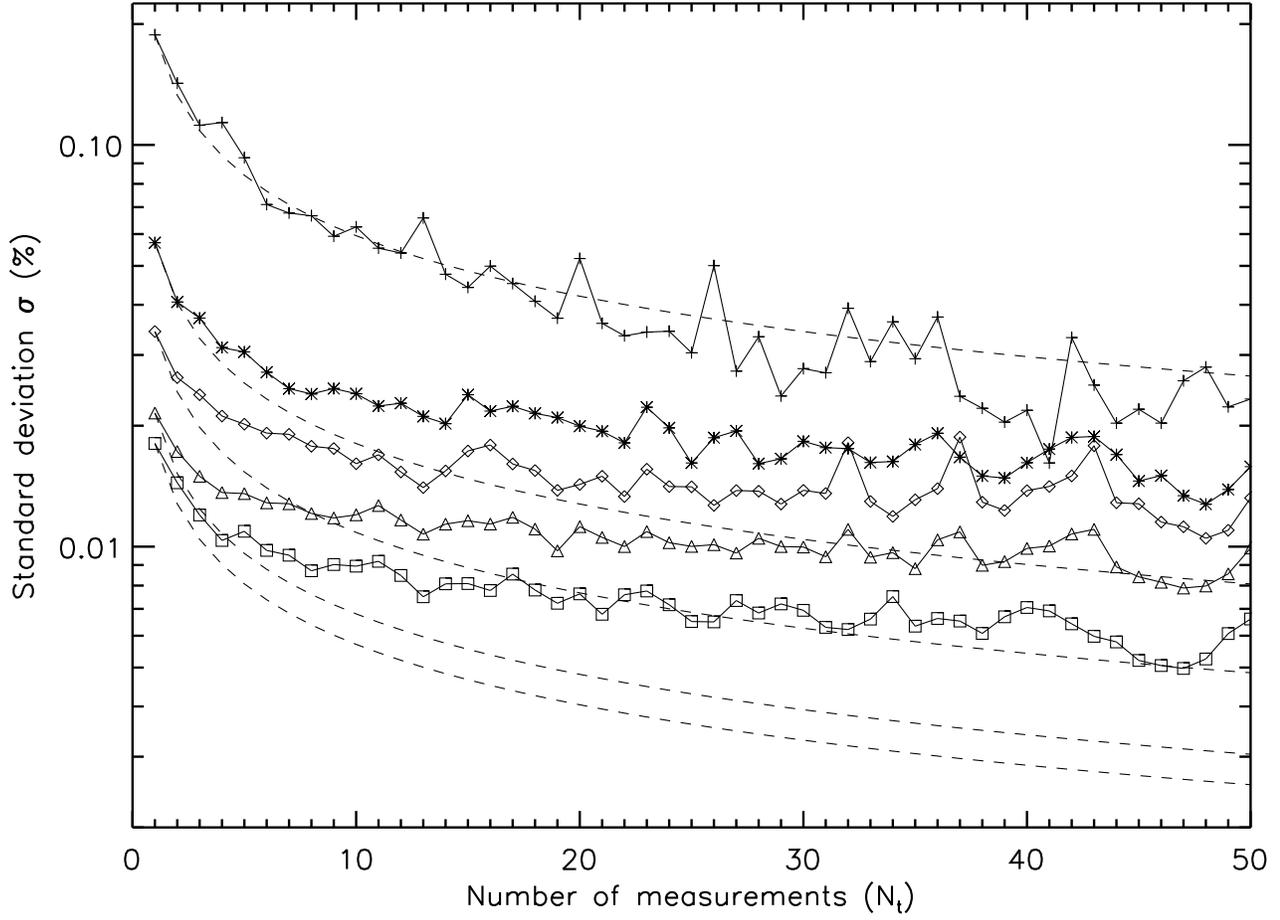}
  \figcaption{Red nose estimation.  Plotted is
  the  standard   deviation  of   the  average  of   N$_t$  successive
  measurements  centered at 6200  {\AA}, binning  in wavelength  by: 1
  pixel (4.8 {\AA},  crosses); 10 pixels (49 {\AA},  stars); 50 pixels
  (244 {\AA}, diamonds); 150  pixels ($\sim$732 {\AA}), and 200 pixels
  (977  {\AA},  squares).   The  dotted  line  shows  an  uncorrelated
  $\sigma$$N_{t}^{-1/2}$  relation.  A  single pixel is  dominated by
  photon noise and follows an uncorrelated relation.  Correlated noise
  appears   when   binning  by   wavelength   pushes  $\sigma$   below
  $\sim$0.02\%.  In this  case, we are able to  reach precision levels
  of   $\sim$0.006\%  (S/N   $\sim$16,000)  when   binning   by  $>$30
  measurements over  $\sim$1,000 {\AA} where we are  dominated by residual
  correlated noise resulting from systematic errors.}
\vspace{0.4cm}
\end{figure}

\begin{figure*}
  \plotone{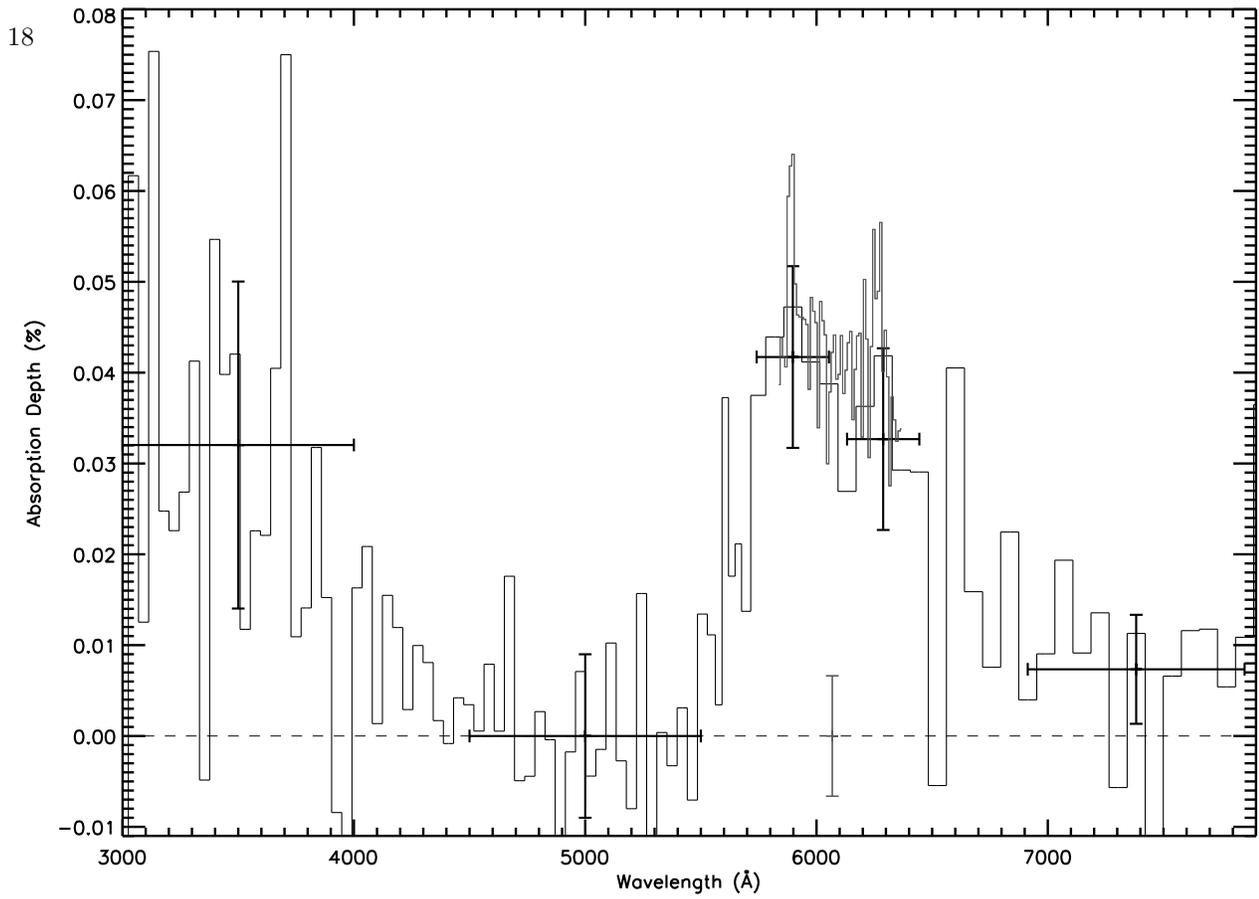}
  \figcaption{Optical   transit  transmission  spectra  of
  HD209458b.  Plotted is the transmission spectra derived from the low
resolution G430L and  G750L gratings (thin black line;  binned over 16
pixels) as  well as the medium  resolution G750M (grey; binned over 18
pixels, $\sim$10{\AA}). Also plotted are 1$\sigma$ uncertainty levels (y-axis error bars) on
low resolution broadband features  averaged over selected wavelength
regions  (x-axis  error bars) and a representative error bar for the binned medium resolution data (grey error bar).   The absorption scale has been normalized to the
value at 5,000{\AA} (1.444\%).  The  uncertainty levels include the effects of red noise.}
\end{figure*}

\end{document}